\documentclass[prd,preprintnumbers,
eqsecnum,floatfix,letterpaper,superscriptaddress,nofootinbib]{revtex4}
\usepackage{color,soul}
\usepackage{amsmath,amssymb,graphicx
}
\usepackage{bm}
\usepackage{times}
\usepackage{tabularx, booktabs,fixme}
\usepackage{microtype}
\usepackage{enumitem}
\usepackage{booktabs}
\usepackage{subfigure}
\usepackage{verbatim}
\usepackage[normalem]{ulem}
\usepackage[varg]{txfonts}
\usepackage[colorlinks, pdfborder={0 0 0}]{hyperref}
\newcolumntype{Y}{>{\centering\arraybackslash}X}
\definecolor{LinkColor}{rgb}{0.75, 0, 0}
\definecolor{CiteColor}{rgb}{0, 0.5, 0.5}
\definecolor{UrlColor}{rgb}{0, 0, 0.75}
\hypersetup{linkcolor=LinkColor}
\hypersetup{citecolor=CiteColor}
\hypersetup{urlcolor=UrlColor}
\allowdisplaybreaks
\newcommand{\FigStart}{\begin{figure}[h]}
\newcommand{\FigEnd}{\end{figure}}

\newcommand{\imrphenb}{\texttt{IMRPhenomB}}	% don't add space here! 
\begin{document}
\title{Projected constraints on the dispersion of gravitational waves
using advanced ground- and space-based interferometers}
\date{\today}
\author{Anuradha Samajdar}
\email{anuradha1115@iiserkol.ac.in }
\affiliation{IISER-Kolkata, Mohanpur, West Bengal 741252, India}
\author{K. G. Arun}
\email{kgarun@cmi.ac.in}
\affiliation{Chennai Mathematical Institute, Siruseri, 603103 India}
\affiliation{Institute for Gravitation and the Cosmos, Pennsylvania  State University, State College, PA 16802}
\begin{abstract}
Certain alternative theories of gravity predict that gravitational waves will disperse as they travel from
the source to the observer. The recent binary black hole observations by Advanced-LIGO have set limits on
a modified dispersion relation from the constraints on their effects on gravitational-wave propagation.
Using an identical modified dispersion, of the form $E^2=p^2c^2+{\mathbb A}\; p^{\alpha} c^{\alpha}$, 
where ${\mathbb A}$ denotes the magnitude
of dispersion and $E$ and $p$ are the energy and momentum of the gravitational wave, we estimate the
projected constraints on the modified dispersion from observations of compact binary mergers by third-
generation ground-based detectors such as the Einstein Telescope and Cosmic Explorer as well as the
space-based detector Laser Interferometer Space Antenna. We find that third-generation detectors would
bound dispersion of gravitational waves much better than their second-generation counterparts. The Laser
Interferometer Space Antenna, with its extremely good low-frequency sensitivity, would place stronger
constraints than the ground-based detectors for $\alpha \leq 1$, whereas for $\alpha > 1$, the bounds are weaker. We also
study the effect of the spins of the compact binary constituents on the bounds.
\end{abstract}
\preprint{}
\maketitle
\section{Introduction}
The direct detection of gravitational waves (GWs) by the LIGO and Virgo
collaborations~\cite{gw150914,gw151226,O1BBH,gw170104} are giving us the
first glimpses of the strong-field dynamics associated with the
mergers of binary black holes.  We now have the first constraints on the deviation
from the post-Newtonian
coefficients~\cite{Arun-test-pn1,Arun-test-pn2,CKM2010,ppe,Li2012,Agathos2014,O1BBH,gw170104,AbbottTGR},
mass of the graviton~\cite{Will98,AW09,O1BBH,gw170104,AbbottTGR}, and consistency
between inspiral and merger-ringdown phases of the binary
evolution~\cite{Ghosh2016,O1BBH,AbbottTGR,gw170104}. The latest addition to the set of
tests is the constraint on the possible dispersion of gravitational
waves~\cite{MYW11,YYP16,gw170104}. If the
propagating GWs disperse, then the dispersion will lead to dephasing of the
GW signal~\cite{MYW11,YYP16}. The consistency of the observed phase
with that of general relativity (GR) will hence set limits on possible dispersion. 
The results for the constraints on modified
dispersion, from the three binary black hole (BBH) detections,  are presented in Fig. 5 of
Ref.\cite{gw170104}. While these bounds are the first from the gravity sector for superluminal
propagation of GWs, the bounds from gravitational Cherenkov radiation
(though very much model dependent) are better than these for subluminal
propagation~\cite{KY15,Kostelecky15,Tasson2016}.

 One natural way to
invoke dispersion of GWs is to postulate that the underlying theory of gravity
does not respect local Lorentz invariance, one of the
fundamental pillars of GR. Hence, the bounds on dispersion can be
translated to constraints on parameters of Lorentz violating theories of
gravity~\cite{YYP16}. Using  GW150914, the
first BBH detected by LIGO, 
Ref.~\cite{SME-LIV2016} discusses constraints on the Standard Model
extension, a generic framework to model Lorentz violating theories of
gravity~\cite{SME,DataTable}. The accuracy on the delay time between the two LIGO
detectors was used to constrain the speed of GW using
GW150914 in Refs.~\cite{Blas2016,Cornish2017}. Using the inferred parameters and constraints on the
post-Newtonian phasing coefficients of GW150914 and
GW151226~\cite{AbbottTGR, O1BBH},
Ref.~\cite{YYP16} discusses the bounds on possible Lorentz violation.

 Improved sensitivities of next generation ground- and
space-based detectors can significantly improve these bounds, possibly
ruling out certain classes of alternative theories of gravity which predict
dispersion of GWs. This forms the theme of this paper in which we obtain the
projected bounds on constraining modified dispersion of GWs using third-generation (3G)
ground-based detectors such as Einstein Telescope (ET)~\cite{etPuntoro}
and Cosmic Explorer (CE)~\cite{CEpsd} as well as the space-based detector 
Laser Interferometer Space Antenna LISA~\cite{KD96},
expected to be launched in the early 2030s, preparations for which are underway.

 Such investigations have been carried out in the past by several
authors. Following the proposal by Will~\cite{Will98}, constraints on
the mass of the graviton using advanced ground- and space-based detectors
were studied in Refs.~\cite{Will98,BBW05a,AW09} using post-Newtonian
gravitational waveforms that account for the inspiral phase of the
binary evolution. Using analytical waveforms that extend beyond
inspiral and account for the merger and ringdown of the binary, Keppel and Ajith~\cite{KA10} 
carried out a similar study for the bounds on graviton mass using
advanced GW detectors for nonspinning systems. Mirshekari \emph{et al.}~\cite{MYW11} proposed an
extension of this idea to include dispersion relations that include
Lorentz violation (which is what we follow here) and deduced the bounds on the modified dispersion
using nonspinning post-Newtonian waveforms~\cite{Bliving,BDEI04}. Reference~\cite{Hansen2014} discussed 
bounds on Lorentz violating theories of gravity using GW observations with and without
electromagnetic counterparts.

 A recent study by Chamberlain and Yunes~\cite{CY17} made a
detailed analysis of the improvement on the constraints on 
several alternative theories of gravity from advanced ground- and space-based GW detectors.  
Using the parametrized post-Einsteinian formalism~\cite{ppe}, they studied systems similar to the
gravitational signal GW150914 as well as other canonical binary black
holes for ground-based detectors and binaries involving supermassive black holes
for space-based detectors. Amongst others, the alternatives include 
the presence of a massive graviton in the GW dispersion relation 
and specific Lorentz violating theories, namely, the
Einstein-Aether and the khronometric theories.
They used propagation effects at the first post-Newtonian order to derive
the mass of the graviton, while correction to the Newtonian GW
phasing was used to put bounds on certain Lorentz
violating theories of gravity. 

 In this work, we consider more realistic
waveforms that account for inspiral, merger, and ringdown phases
as well as
study the effect of the presence of spins. We extend the
analysis to include generic dispersion and derive bounds on the magnitude of
dispersion for different types of modifications and for different detector
sensitivities. Our goal here is to discuss the ability of
advanced detectors to constrain the possible dispersion of
GWs without referring to any particular theory of gravity.
We consider only propagation effects here as our aim is to probe
dispersion.

The rest of the paper is organized as follows. 
Section~\ref{sec:basics} describes the modified dispersion relation and the
expression for a resulting dephasing of the gravitational waves.
We introduce the waveform, detector sensitivities, and Fisher matrix
formalism in Sec.~\ref{sec:liv-model}.
The results and conclusions are discussed in
Sec.~\ref{sec:liv-fm-res}, and conclusions and outlook are presented in
Sec.~\ref{sec:liv-fm-conc}.

\section{Constraining Dispersion of gravitational waves}\label{sec:basics}
Following Refs.~\cite{MYW11,YYP16}, we consider a modified dispersion relation
for GWs, which is given by
\begin{equation}
E^2=p^2 c^2 + {\mathbb A}\,p^{\alpha}c^{\alpha},\label{eq:MDR}
\end{equation}
where $E$ and $p$ are the energy and momentum of GWs and ${\mathbb A}$ denotes the
magnitude of dispersion corresponding to the exponent $\alpha$. As shown in
Ref.~\cite{MYW11}, this modified dispersion relation leads to a dephasing of the
gravitational signal given, in the frequency domain, by
\begin{equation}
\Psi_{\mathrm{total}} (f) = 
\begin{cases} \Psi_{\rm GR}(f) - \zeta
u^{\alpha-1} \;\;\; \alpha \ne 1, \\ 
\Psi_{\rm GR}(f) + \zeta \ln{u} \;\;\;\alpha = 1.
\end{cases}\label{eq:phase} 
\end{equation}
In the above, $u = (\pi \mathcal{M}
f)$, where $\mathcal{M}$ is the chirp mass of the binary and $f$ is the
GW frequency. $\zeta$ is given by
\begin{equation} 
\zeta = 
\begin{cases} 
\frac{\pi^{2-\alpha}}{(1-\alpha)}\frac{D_\alpha}{\lambda_\mathbb{A}^{2-\alpha}}\frac{\mathcal{M}^{1-\alpha}}{(1+z)^{1-\alpha}}
\;\;\;  \alpha \ne 1, \\ 
\frac{\pi D_1}{\lambda_\mathbb{A}} \;\;\;\alpha = 1, 
\end{cases} \label{eq:zeta}
\end{equation}
 
where $\lambda_\mathbb{A} \equiv h c \mathbb{A}^{\frac{1}{\alpha-2}}$ (with $c$ and $h$
referring to the speed of light and Planck constant, respectively)
denotes the length scale introduced by the dispersion and
\begin{equation}
D_{\alpha}=\frac{(1+z)^{1-\alpha}}{H_0}\int_0^z\;\frac{(1+z')^{\alpha-2}}{\sqrt{\Omega_m(1+z')^3+\Omega_{\Lambda}}}
dz',
\end{equation}
is a distance measure introduced by dispersion. 

$\Omega_m$ and $\Omega_\Lambda$ are, respectively, the matter and dark
energy fractions in a (flat) $\Lambda_{\rm CDM}$ model of cosmology, 
for which we use the values (0.3065, 0.6935) estimated by the Planck Collaboration~\cite{Planck}.

The group velocity of GWs, with the modified dispersion relation, can
easily be obtained by differentiating it with respect to $p$, which to
the leading order in ${\mathbb A}E^{\alpha-2}$, reads
%\begin{equation}
$v_{\rm gw}=c\left(1+\frac{\alpha-1}{2} {\mathbb
A}\,E^{\alpha-2}\right)$.
%\end{equation}
Depending on the sign of $\mathbb{A}$ and the value of $\alpha$, GWs may propagate superluminally or
subluminally. The bounds from GW observations, reported in
Ref.\cite{gw170104}, have been derived for both these sectors (see Fig.~5 of
Ref.\cite{gw170104}). However, using the parameter estimation method that we employ here, 
we cannot obtain bounds for these two sectors separately.

Since the method explored here is generic, $\alpha$ can take any
value greater than or equal to 0, depending on the alternative theory.
Here, we consider the
representative cases of $\alpha=0, 1, 2.5, 3$. The $\alpha=0$ bounds can
easily be mapped onto a bound on graviton mass (assuming $\mathbb{A} >0$). 
The $\alpha=1$ modification, as can be seen from
Eq.~(\ref{eq:zeta}), is a special case that brings in logarithmic
correction to the GR phasing. Modifications with $\alpha=2.5$
and $\alpha=3$ correspond to certain Lorentz violating alternative theories
of gravity such as multifractal spacetime~\cite{multifractal} and
doubly special relativity~\cite{DSR1}, respectively.

The goal of this paper is to calculate the projected accuracy with
which the magnitude of dispersion parameter ${\mathbb A}$ can be bounded by future
observations of compact binaries by advanced ground- and
space-based detectors, as a function of the total mass of the compact
binaries for different values of $\alpha$.  Dimensionally,
${\mathbb A}$ (for a given
$\alpha$) has the unit of ${\rm energy}^{2-\alpha}$, and hence our bounds
are reported in units of ${\rm eV}^{2-\alpha}$, a convenient unit for all $\alpha$.
 These bounds are obtained by using the expected sensitivities of the
future GW detectors and using the parameter estimation technique of
Fisher information matrix where the compact binary waveforms will be
modeled by the \imrphenb\  model restricting to equal-mass binary black
hole mergers (which will be representative of the typical bounds even
for asymmetric binaries).

\begin{figure}[t!] \centering
\subfigure{
\includegraphics[keepaspectratio,width=0.45\textwidth]{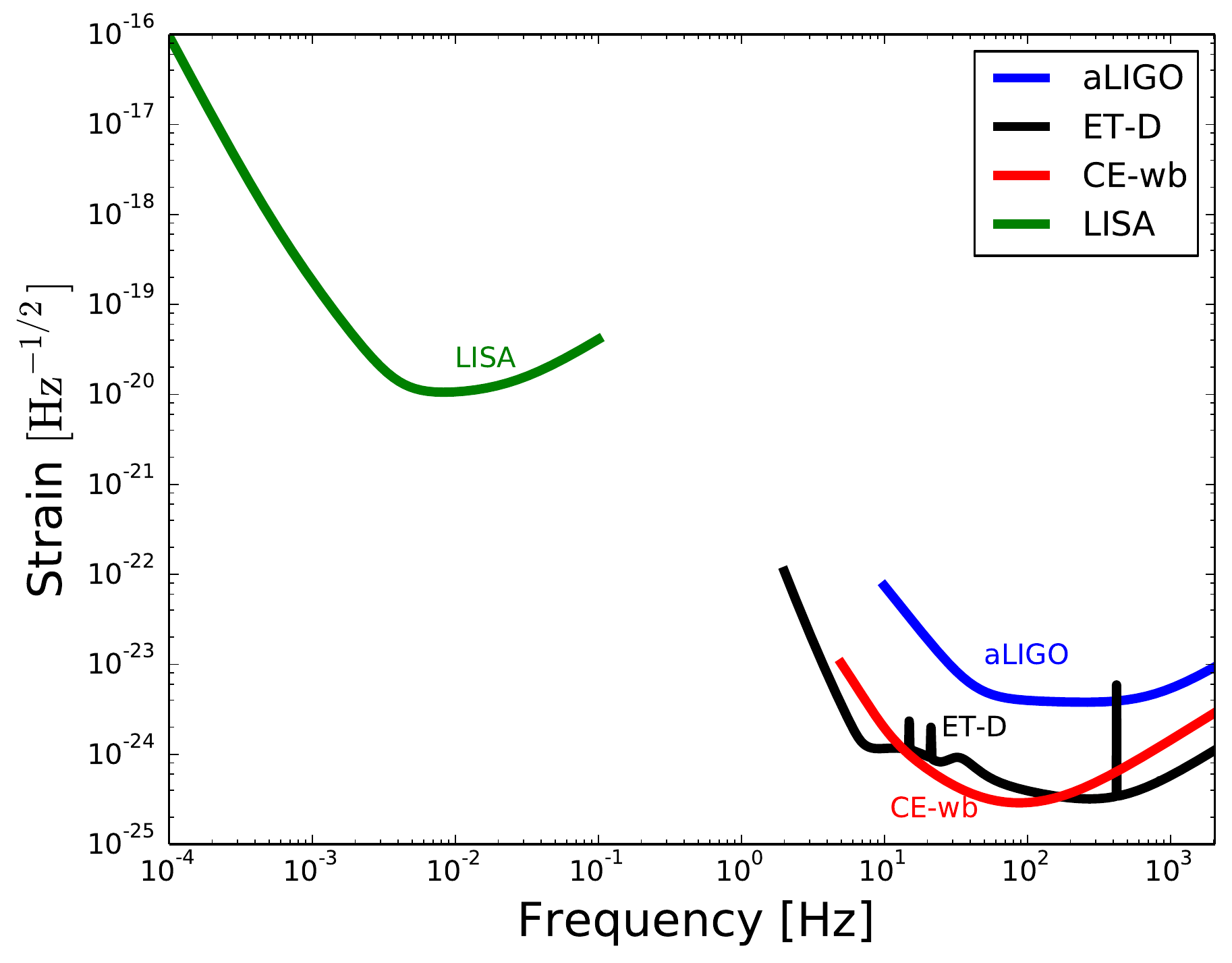}}
\includegraphics[keepaspectratio,width=0.45\textwidth]{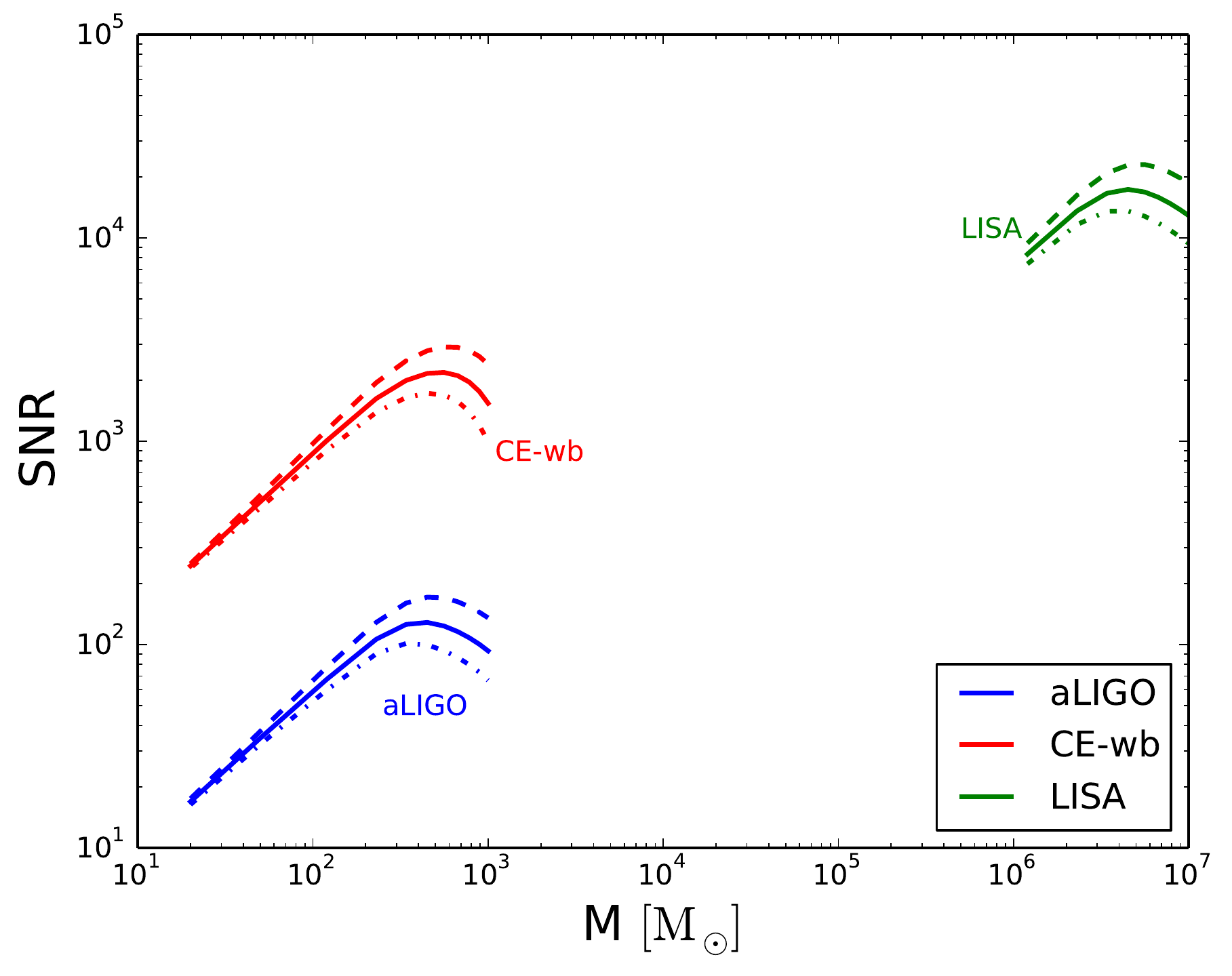}
\caption{Strain sensitivities of ET-D (black curve), CE-wb (red curve),
\texttt{aLIGOZeroDetHighPower} (blue curve), and LISA (green curve)
overlaid. The corresponding signal-to-noise ratios as a function of the
total mass of the compact binary are on the right panel. The sources
considered consist of nonspinning (solid curves), with
$\chi=0.4$ (dashed curves) and with $\chi=-0.4$ (dash-dotted curves).
The sources targeted by CE, aLIGO, and ET are at a redshift of 0.2, and the 
LISA sources are at a redshift of 0.5.} \label{fig:3g-psds}
\end{figure} 
\section{Analysis Set up}\label{sec:liv-model}
\subsection{Waveform model}\label{sec:liv-3g-imr}
We use the analytical waveform model
\imrphenb\ ~\cite{AjithPhenB} as an example of a gravitational waveform containing
the inspiral, merger, and ringdown phases of the binary's
evolution.
This waveform family is obtained by combining the post-Newtonian
description~\cite{Bliving,BDEI04, ABFO08} of the inspiral with a set of
numerical relativity simulations (up to a mass ratio of 4) accounting for 
spin effects when the spins are (anti)aligned with respect to the orbital
angular momentum vector of the binary. A more recent 
family of waveforms, \texttt{IMRPhenomD}~\cite{imrpv2_2}, 
is calibrated to 
numerical simulations with higher mass ratios up to 18. However,
we focus on equal-mass systems for which the two waveforms do not differ significantly.
The definition of $\chi$ in the above equation
depends on the two component masses $m_1$ and $m_2$ and the corresponding
dimensionless spin parameters $\chi_1$ and $\chi_2$ where
$\chi_1\equiv\frac{|\vec{S}_1|}{m_1^2}$ and similarly for $\chi_2$.
Schematically, the waveform reads
\begin{equation}
{\tilde h}(f)={\cal C}\; {\cal B}(f;
M, \eta, \chi)\,e^{i\,\Psi(f; M, \eta, \chi)},
\end{equation}
where $M$ is the total mass, $\eta$ the symmetric mass ratio and $\chi$ is
the effective spin parameter. ${\cal C}$ encodes information about the
luminosity distance, source location, and orientation, whereas ${\cal B}$
contains the dependences on the intrinsic parameters (masses and spins). 
The exact waveform we use is given in Eq.~1 and Table I of
Ref.\cite{AjithPhenB}. The waveform is truncated at the frequency referred
to as $f_3$ (and given in Table 1) of Ref.\cite{AjithPhenB}.
The phase of the \imrphenb\ waveform is deformed
accounting for GW dispersion following Eq.~(\ref{eq:phase}).

\subsection{Sensitivity of future detectors} \label{sec:liv-fm-dets}
The detector noise is modeled as a stationary, zero-mean Gaussian, random process. The 
assumption of stationarity implies that the noise properties do not change over time.
If ${\tilde n}(f)$ is the Fourier transform of the noise $n(t)$, the
noise power spectral density (PSD) $S_h(f)$ is defined by
\begin{equation}
 \langle \tilde{n}(f) \tilde{n}^*(f') \rangle = \frac{1}{2}S_h(f) \delta
(f-f'),
\end{equation}
where $\delta$ denotes the Dirac delta function.
In this section, we list the sensitivities of different detector configurations
we use in the present work: the ET, CE, and LISA.
For comparison of results, we also list the design sensitivity of advanced LIGO
detector.
\subsubsection{Design sensitivity of AdvLIGO} 
An analytic fit to Advanced LIGO's zero-detuned-high-power (called
\texttt{aLIGOZeroDetHighPower}) PSD is given in
Ref.~\cite{Ajith11} as
\begin{equation} S_h(f) =
10^{-48} \left( 0.0152 x^{-4} + 0.2935 x^{9/4} + 2.7951 x^{3/2} - 6.5080 x^{3/4}
+ 17.7622 \right)\ \mathrm{Hz}^{-1}, \label{eq:aZDHP}
\end{equation} where $x=f/245.4$.
For the studies done with Advanced LIGO sensitivity below, we use a lower-frequency cutoff of 10 Hz.
\subsubsection{Einstein Telescope} 
ET is an envisaged 3G detector with proposed
frequency sensitivity in the range of $1-10^4$ Hz. 
Details of its
sensitivity design are given by Hild \emph{et al.}~\cite{Hild-etal-2011}. In the following study, we use the
sensitivity of the ET-D configuration given in  Ref.~\cite{dcc-doc} 
with a 2 Hz lower-frequency cutoff.

\subsubsection{Cosmic Explorer} Dywer \emph{et al.}~\cite{Dwyer-CE}
introduced the idea of a ground-based interferometer with an arm length
of 40 km, which is referred to as CE. It has been argued that 40
km is the optimal arm length beyond which no additional scientific gain would be
evident. 
Various noise sources
corresponding to CE are also discussed by Abbott \emph{et al.}~\cite{CEpsd},
from which we have used an
analytical fit to the CE-wb configuration~\cite{Sathya2016} given by
 \begin{equation} S_h(f) =
10^{-50} (11.5 f_{10}^{-50} + f_{25}^{-10} + f_{53}^{-4} + 2 f_{80}^{-2} + 5 + 2
f_{100}^2)\ \mathrm{Hz}^{-1}, \label{eq:psd-ce} \end{equation}
where $f_k \equiv (f/k)$ Hz. In the following discussion,
we shall mean the CE-wb configuration when we refer to the CE sensitivity. 
We use a low-frequency cutoff of 5 Hz for our studies below.

\subsubsection{LISA}
LISA was proposed as a space-based GW
observatory sensitive to a frequency range $\sim 10^{-4}-0.1$ Hz and
capable of observing mergers of supermassive binary black holes with
masses between $\sim 10^4-10^7 \ \mathrm{M}_\odot$. There
is increased enthusiasm about LISA after the promising scientific output from LISA Pathfinder~\cite{LPF2016}.
We use the latest
noise PSD of LISA used by Babak \emph{et al.}~\cite{Babak17} given by \begin{equation}
S_h(f) = \frac{20}{3} \frac{4S_n^{acc}(f) + 2S_n^{loc} + S_n^{sn} +
S_n^{omn}}{L^2} \times \left[ 1+\left( \frac{2Lf}{0.41c} \right)^2 \right]\
 \mathrm{Hz}^{-1}, \end{equation} where $L$ is the arm length, now
considered to be $2.5 \times 10^9$ m, $S_n^{acc}(f)$, $S_n^{loc}$, $S_n^{sn}$, and
$S_n^{omn}$ are, respectively, the noise contributions due to the low-frequency
acceleration, local interferometer noise, shot noise, and other measurement
noise. The low-frequency noise is given by \begin{equation} S_n^{acc}(f) =
\left\{ 9 \times 10^{-30} + 3.24 \times
10^{-28}\left[\left(\frac{3\times10^{-5}\ \mathrm{Hz}}{f}\right)^{10}
+\left(\frac{10^{-4}\ \mathrm{Hz}}{f}\right)^2\right]  \right\} \left(\frac{1 \
\mathrm{Hz}}{2\pi f}\right)^4 \ \mathrm{m}^2 \ \mathrm{Hz}^{-1}.  \end{equation}
The other noise components are given by \begin{equation} \begin{split} S_n^{loc}
&= 2.89 \times 10^{-24} \ \mathrm{m}^2 \ \mathrm{Hz}^{-1}, \\ S_n^{sn} &= 7.92
\times 10^{-23} \ \mathrm{m}^2 \ \mathrm{Hz}^{-1}, \\ S_n^{omn} &= 4.00 \times
10^{-24} \ \mathrm{m}^2 \ \mathrm{Hz}^{-1}. \\ \end{split} \end{equation}

Figure~\ref{fig:3g-psds} shows the sensitivities of all configurations of the
detectors used here. 

\subsection{Fisher Information Matrix}\label{sec:Fisher}
\begin{figure}[t!]
\includegraphics[keepaspectratio,width=0.6\textwidth]{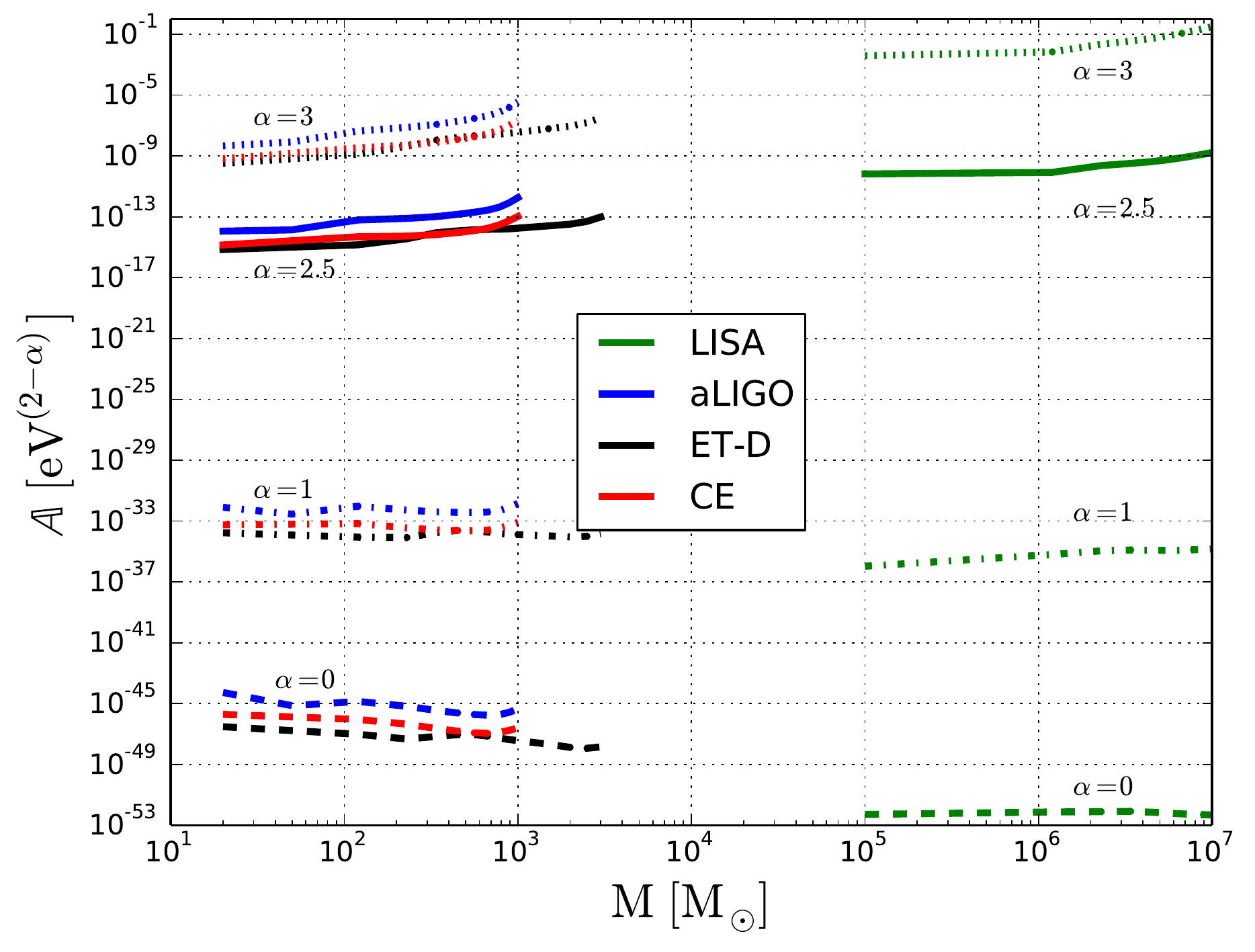}
\caption{Upper bounds on $\mathbb{A}$ obtained with equal-mass nonspinning
sources for future ground-based detectors and the planned space-based detector
LISA. For ground-based detectors, we use the future design of aLIGO (in blue),
the 3G detectors ET (in black) and CE (in red). The total masses vary from
$20-1000\ \mathrm{M}_\odot$ for aLIGO and CE and lie between $20-3000\ \mathrm{M}_\odot$
for ET.
The total masses
vary between $10^5-10^7\ \mathrm{M}_\odot$ for sources targeted by LISA (in
green). All bounds are obtained for $\alpha=0$ (dashed lines), $\alpha=1$
(dash-dotted lines), $\alpha=2.5$ (solid lines) and $\alpha=3$ (dotted lines).
The sources targeted by the ground-based detectors are at a redshift
of 0.2 and the LISA sources occur at a redshift of 0.5.}
\label{fig:A-noSpin} 
\end{figure} 

All analyses carried out here
have been done with a Fisher matrix approach~\cite{CF94}. 
Assuming the noise in a GW detector to be Gaussian, the likelihood is
given by  
\begin{equation} p(d|\vec{\theta}) = \exp{\left[ -\frac{1}{2} \Gamma_{ab} \Delta
\theta^a \Delta \theta^b \right]}, \label{eq:fm-lhood}
 \end{equation}
where
$\Gamma_{ab}$ denotes the Fisher information matrix and $\Delta \theta^a$
represents the error in estimation of the parameter $\theta^a$.

$\Delta \theta^a$ is given by $\sqrt{\Sigma^{aa}}$, where $\Sigma$ is
the covariance matrix given by the inverse of the Fisher information matrix
$\Gamma$. The diagonal elements of $\Sigma$ represent the errors whereas the off-diagonal
elements give us the correlation coefficients between the parameters.
Components of the Fisher matrix are given by \begin{equation} \Gamma_{ab} =
\left(\frac{\partial h}{\partial \theta^a}\Bigg|\frac{\partial h}{\partial
\theta^b}\right),  \label{eq:fisher} \end{equation} 
where $h$ is the waveform in frequency domain.
The scalar product notation between two frequency domain waveforms $h_1$ and $h_2$ is defined as
\begin{equation}
 (h_1|h_2) = 2 \mathcal{R} \int_{f_{\rm low}}^{f_{high}} \frac{h_1^*(f) h_2(f) + h_1(f) h_2^*(f)}{S_h(f)} df,
\end{equation}
where $S_h(f)$ is the PSD of the detector. 
The integration in the above is carried out
between a lower cutoff frequency corresponding to the detector and the upper
cutoff frequency, which is the frequency at which the signal terminates.

 For all the
ground-based detectors the  upper frequency cutoff is minimum of the
waveform's termination frequency ($f_3$) as given in Ref.\cite{AjithPhenB},
whereas for LISA, it is ${\rm min}\;(f_3,0.1\,{\rm Hz})$. We have not
considered here the orbital motion of LISA and have instead treated LISA like a static
detector. The orbital motion and the corresponding
modulations to the waveform are likely to be more important for distance
estimation and source localization which are not relevant to the present
analysis. 
However, we note that the orbital motion of the detector would indeed be important for 
detection of GW signals.
For both ground- and space-based detectors, we use only
single detector configurations for our analysis. 

Details of Fisher matrix implementation can be found in
Refs~\cite{CF94,PW95}. 
The errors
computed from the Fisher matrix are a lower bound on the actual errors when
the signal-to-noise ratio (SNR) is high and the noise is Gaussian. 
Since these
assumptions are likely to be  hold,  as can be seen from the right
panel of Fig.~\ref{fig:3g-psds}, for most detections using advanced detectors,
we believe Fisher matrix-based estimates would suffice here. One may refer to
Ref.~\cite{Vallisneri07} for a detailed discussion on the domain of
applicability of the Fisher matrix.
\begin{figure}[t!] \centering
\mbox{\subfigure{\includegraphics[keepaspectratio,width=0.5\textwidth]{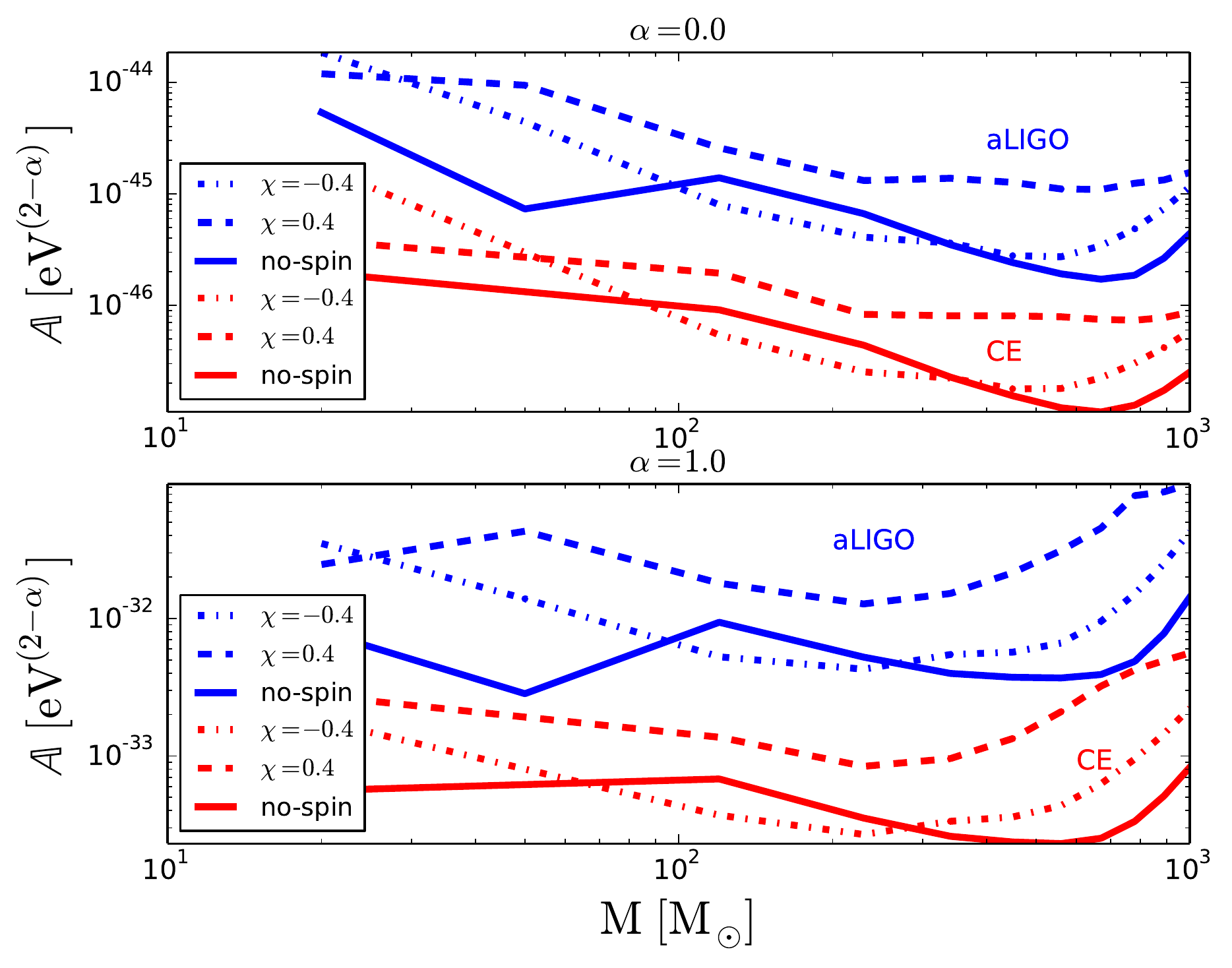}}\quad
\subfigure{ \hspace{-0.2cm}
\includegraphics[keepaspectratio,width=0.5\textwidth]{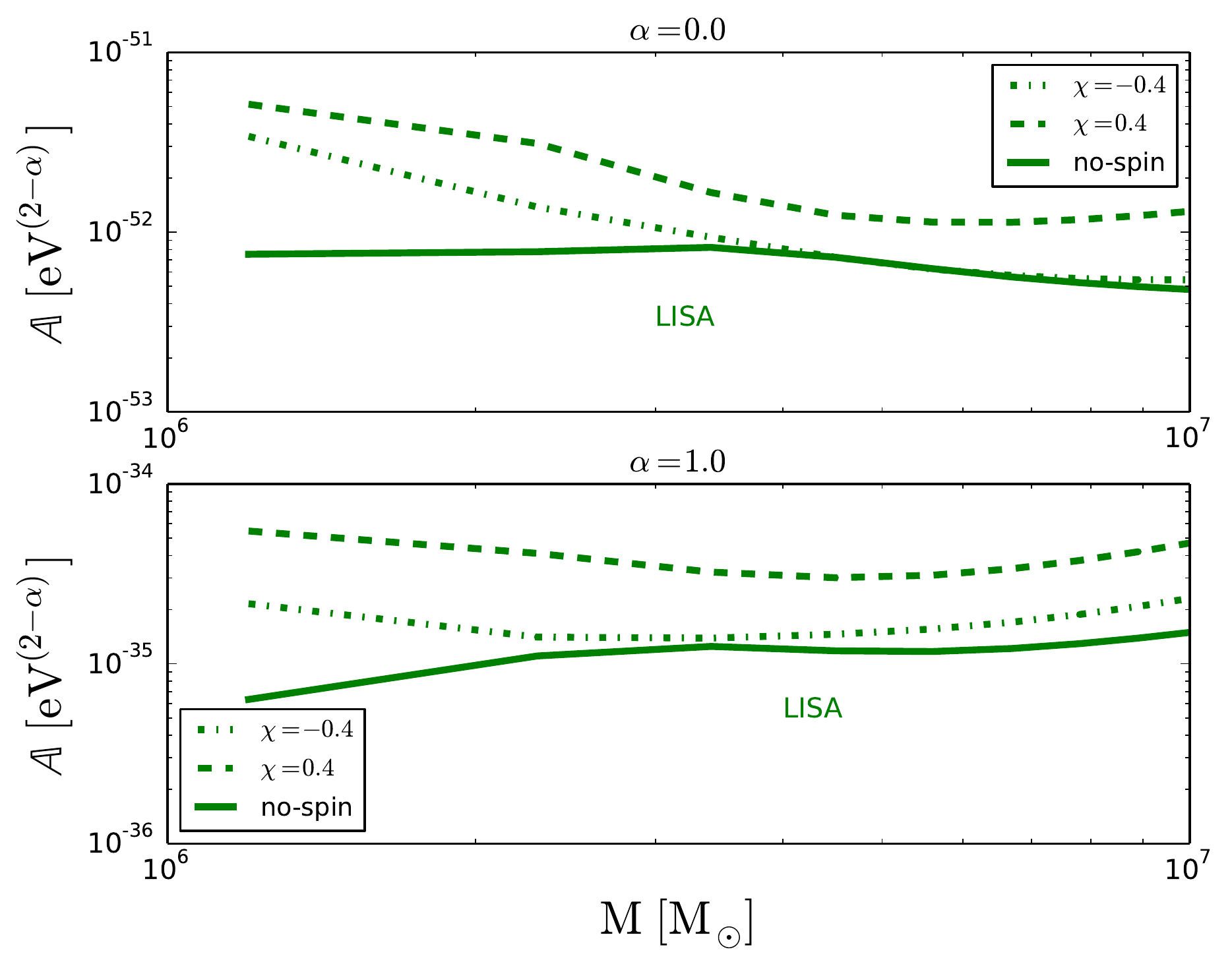}
}} \caption{Upper bounds on $\mathbb{A}$ varying with total mass for spinning
sources overlaid with nonspinning sources for CE and aLIGO (left) and LISA
(right). The bounds are for $\alpha=0$ (top panel) and $\alpha=1$ (bottom
panel). The plots are made for $\chi=0.4$ (dashed lines), $\chi=-0.4$ (dash-dotted
lines) and non-spinning sources.  In general, spins deteriorate the bounds,
though $\chi=-0.4$ is found to perform almost comparably with the nonspinning
counterparts for higher mass sources. It is discussed in
Sec.~\ref{sec:liv-fm-spin}.} \label{fig:A-spin-alpha-lt2} 
\end{figure}

\section{Calculation of the bounds on dispersion} \label{sec:liv-fm-res} 
For our studies, we use
equal-mass systems at a distance of 1 Gpc ($z\simeq 0.2$)
for the ground-based detectors. The right-hand panel of Fig.~\ref{fig:3g-psds} shows a comparison
of the SNRs from the future detectors. Henceforth, we shall use aLIGO to mean
the improved zero-detuned-high-power \texttt{aLIGOZeroDetHighPower}
sensitivity.
For the space-based detector LISA, we use equal-mass systems 
located at a distance
of 3 Gpc ($z\simeq 0.5$). We have reproduced very closely the results of
Ref.\cite{KA10} for the $\alpha=0$ case and of Ref.\cite{MYW11} for the
corresponding $\alpha$ (using the waveform model of Ref.\cite{MYW11}). 
Using the model of modified dispersion described in
in the Introduction, the waveform model of Sec.~\ref{sec:liv-3g-imr}, and
sensitivities of advanced detectors in Sec.~\ref{sec:liv-fm-dets}, we
compute the errors on $\zeta(\alpha)$ for different $\alpha$ values and
convert the errors $\Delta \zeta$ to upper bounds on ${\mathbb A}$ using
the expression for $\zeta$. We compare the bounds obtained with and
without the inclusion of spins in the parameter space in the next two
subsections. 

\subsection{Bounds from nonspinning sources}\label{sec:liv-fm-nonspin} 

To derive the bounds on ${\mathbb A}$ for nonspinning binaries, we
use the parameter space given by $\vec{\theta} \equiv \{\log {\cal C},
\phi_c, t_c, \log M, \log \eta, \zeta\}$.
For aLIGO and CE detectors, we use sources with total
masses lying between $20$ and $1000 \ \mathrm{M}_\odot$, and for the ET, we use sources
with total masses lying between $20$ and $3000 \ \mathrm{M}_\odot$. 
For LISA, we use total masses
lying between $10^5$ and $10^7 \ \mathrm{M}_\odot$. These choices are
motivated by the sensitivities of the detectors.
Figure~\ref{fig:A-noSpin} shows the bounds for $\alpha=\{0,1,2.5,3 \}$ as
representative cases of $\alpha<2$ and $\alpha>2$ for the advanced LIGO, ET, CE, and LISA detectors.
In terms of broad features, one finds that as we increase $\alpha$ from
0 to 4 the bound on ${\mathbb A}$ worsens very rapidly by about 54
orders of magnitude for ground-based detectors and 58 orders of magnitude
for LISA. This has been known in the
literature~\cite{MYW11,YYP16,gw170104} and can be attributed to the fact
that higher $\alpha$ induce phase corrections at higher frequencies
(higher post-Newtonian orders, if one naively views the phase
corrections to be post-Newtonian-like). Since gravitational-wave detectors have less capability to
constrain phase deformations at higher
orders~\cite{Arun-test-pn2,ppe,Agathos2014}, this is naturally
expected.

We next note that for $\alpha=0$
the phase deformations are degenerate with that due to a mass of the
graviton, for which the upper
bounds on the dispersion parameter $\mathbb{A}$ goes as $\mathbb{A} \equiv m_g^2$.  The
bounds get worse with higher $\alpha$. Among
the ground-based detectors, ET performs better than CE though they
perform comparably at sources with higher mass.
The sensitivity of ET at high and low frequencies is better than
CE, as can be noted from the 
sensitivity plot in Fig.~\ref{fig:3g-psds} which explains why the bounds from ET are better
for lower mass sources than from CE.
They both outperform aLIGO by about an order of magnitude.
  
Bounds from LISA are much better
for $\alpha=0,1$ than those obtained from the other detectors. For $\alpha=0$,
this is what has been observed by Keppel and Ajith~\cite{KA10}.
 However, for $\alpha > 1$, the bounds from  LISA are worse compared to
the ground-based detectors and they become progressively worse as we go
to higher values of $\alpha$. This somewhat unexpected trend may be
explained by noting that the dephasing due to modified dispersion scales
as $\delta \Psi\sim {\mathbb A} f^{\alpha-1}$, and hence for a given
${\mathbb A}$, the dephasing will be larger in the LISA band for $\alpha
\leq 1$ whereas for $\alpha > 1$, dephasing will be larger for the
ground-based detector band ($f\geq 1 Hz$). A larger dephasing would
imply better prospects for constraining the parameter ${\mathbb A}$, as
seen in the figure.

\begin{figure}[t!] 
\centering
\mbox{\subfigure{\includegraphics[keepaspectratio,width=0.5\textwidth]{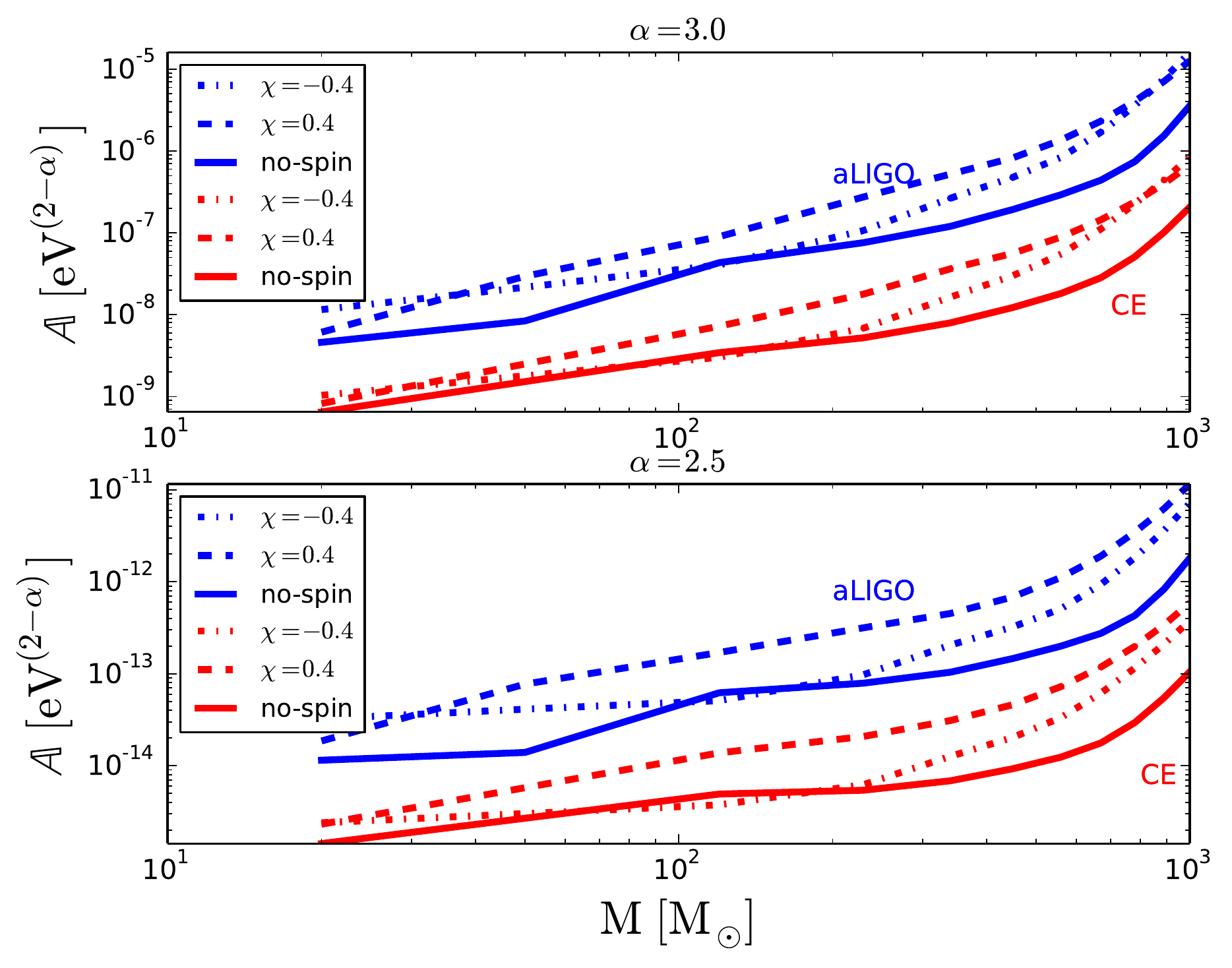}}\quad
\subfigure{ \hspace{-0.2cm}
\includegraphics[keepaspectratio,width=0.5\textwidth]{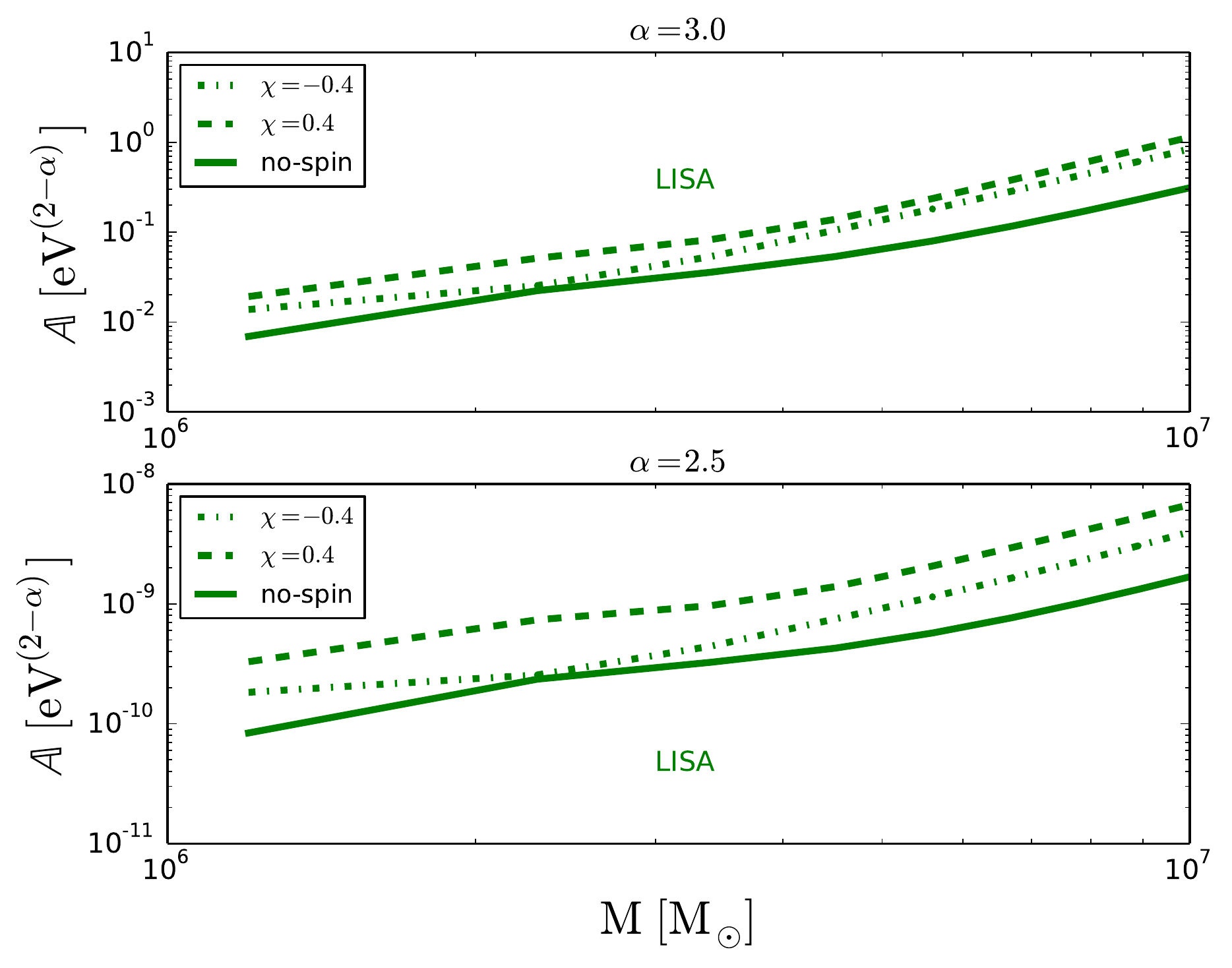}
}} \caption{Upper bounds on $\mathbb{A}$ varying with total mass for spinning
sources overlaid with nonspinning sources for CE and aLIGO (left) and LISA
(right). The bounds are for $\alpha=3$ (top panel) and $\alpha=2.5$ (bottom
panel). The plots are made for $\chi=0.4$ (dashed lines), $\chi=-0.4$ (dash-dotted
lines), and nonspinning sources. A more detailed discussion occurs in
Sec.~\ref{sec:liv-fm-spin}.} \label{fig:A-spin-alpha-gt2} 
\end{figure}

 For $\alpha=0$, we have compared our bounds obtained on $\lambda_\mathbb{A}$
with that of the graviton Compton wavelength $\lambda_g$ reported by Keppel and Ajith~\cite{KA10}.
We have compared our bounds with those reported in Table III in Ref.~\cite{KA10} for equal-mass binaries with 
$f_{\rm low}=10$ Hz for ground-based detectors and $f_{\rm low}=10^{-4}$ Hz for LISA and have found reasonable agreement.
%% starting here: 1st Dec. 2017

\subsection{Bounds from spinning sources} \label{sec:liv-fm-spin}
 For spinning sources, we include $\chi$ in our parameter set. We work with the parameter set
$\vec{\theta} \equiv \{\log {\cal C}, \phi_c, t_c, \log M, \log \eta, \zeta, \chi\}$.
For ground-based detectors, we use sources with total
masses lying between $20$ and $1000 \ \mathrm{M}_\odot$, and for LISA, we use total masses
lying between $10^6$ and $10^7 \ \mathrm{M}_\odot$.
For the spin parameter $\chi$, we use a Gaussian prior with
a mean of 0 and standard deviation of 0.3, while calculating
errors. This is motivated by the fact that in all the observed BBH
mergers so far measured values of $\chi$ are small and close to zero. However we have chosen
the values of $\chi=\pm0.4$ to study the effect of spins and their
alignment, which are greater than the width of the prior so that
we are not  severely limited by the priors.
 We see that the bounds in general worsen with inclusion of spins, as
expected when we add a new parameter without more structure to the
waveform.  The inversion accuracy, defined as largest element in the difference between the identify  matrix and the 
product of the covariance matrix with the Fisher matrix, is $\sim 10^{-3}$ for spinning sources.

Figure~\ref{fig:A-spin-alpha-lt2}
shows a comparison of bounds from the spinning sources and the non-spinning
sources for $\alpha=0,1$. Figure~\ref{fig:A-spin-alpha-gt2} shows the bounds for the same sources at
$\alpha=2.5$ and $3$. 
 
We observe the general trend that systems that have spins antialigned
with respect to the orbital angular momentum yield better bounds than
those of which the spins are aligned with respect to the orbital angular
momentum, despite the SNRs of the former being smaller
than the latter. Since we are measuring a propagation effect, the
bounds are likely to improve when sources are at a larger distance. 
From the right panel of Fig.~\ref{fig:3g-psds},
it is evident that SNRs for the aligned spinning sources are higher than those for the 
antialigned sources. For a fixed source at any $\alpha$, this is as 
if the wave travels a larger effective distance for a negative value of $\chi$.  
The bound is therefore better with a larger propagation distance.
\section{Conclusion and outlook} \label{sec:liv-fm-conc}

As a follow up to the recent LIGO bounds on the dispersion of
GWs~\cite{gw170104}, we extend some of the previous
works~\cite{Hansen2014,CY17} to assess the capabilities of advanced
ground- and space-based interferometers to constrain any possible
dispersion of GWs using binary black hole observations. 
 Our important results are summarized in Table~\ref{tab:summary}, which
presents the typical (median) bounds on dispersion for ground- and space-based
detector configurations, for various types of modification to the
dispersion (different values of $\alpha$). Sources for ground-based detectors are at a
redshift of 0.2 ($\simeq 1 {\rm Gpc}$) whereas those for LISA are at
$z=0.5$ ($\simeq 3 {\rm Gpc}$).
 The numbers in parentheses denote the bounds for
$\chi=0.4$. For $\alpha \leq 1$, the bounds improve by several
orders of magnitude as we go from advanced LIGO to 3G
detectors to LISA. However, for $\alpha >1$, the bounds are worse for
LISA compared to ground-based detectors. In all the cases, 3G
ground-based detectors can constrain GW dispersion much more
stringently than second-generation detectors. As expected, inclusion of spins worsens
the bounds, but the dependence of the bounds on the spins is not
straightforward to understand as the waveform model we employ uses an
effective spin parameter that is a linear combination of masses and
spins.

\acknowledgements
A.S. thanks MHRD for financial assistance. A.S. would like 
to thank Chennai Mathematical Institute for  hospitality during the
initial phase of the project.
K.G.A. acknowledges Grant No. EMR/2016/005594 from Science and Engineering
Research Board (SERB), India. K.G.A. is partially supported
by a grant from Infosys Foundation. K.G.A. acknowledges support from the Indo-US Science and Technology
Forum through the Indo-US Centre for the Exploration of Extreme
Gravity (Grant No. IUSSTF/JC-029/2016). 
We have significantly benefited from discussions from many
members of the LIGO
Scientific Collaboration and Virgo Collaboration. We thank M. Agathos, S. Babak, W. Del
Pozzo, A.Ghosh, C. Mishra, R. Nayak, B. S. Sathyaprakash, C. Van Den
Broeck, S. Vitale for many insightful discussions. K.G.A. thanks L. Stein
and A. Laddha 
 for useful discussions. We thank Archisman Ghosh for critical reading
of the manuscript and much input, which helped us improve the
presentation in the draft.
We thank N. V. Krishnendu for
careful reading of the manuscript. Useful conversations with Stefan
Hild on Einstein Telescope noise PSDs are gratefully acknowledged. 
This research was initiated during the ‘Future of Gravitational
Wave Astronomy Workshop’ at the International Centre for
Theoretical Sciences (code: ICTS/Prog-fgwa/2016/04).

\begin{table*}[t]
\centering
 \begin{tabular}{ c c|c|c|c  } \hline
  \multicolumn{5}{c}{$\mathbb{A}$ [in $\mathrm{eV}^{2-\alpha}$]} \\
 \hline
  \multicolumn{1}{c}{Detector} & $\alpha=0$ & $\alpha=1$ & $\alpha=2.5$ & $\alpha=3$  \\\hline
  aLIGO & $3.50 \times 10^{-46}$ ($1.33\times10^{-45}$) & $4.87\times10^{-33}$ ($3.12\times10^{-32}$) & $1.46\times10^{-13}$ ($6.84\times10^{-13}$)& $1.93\times10^{-7}$ ($8.25\times10^{-7}$) \\
  CE & $1.73\times10^{-47}$ ($8.06\times10^{-47}$) & $3.34\times10^{-34}$ ($2.10\times10^{-33}$) & $1.24\times10^{-14}$ ($7.25\times10^{-14}$) & $1.82\times10^{-8}$ ($8.91\times10^{-8}$)\\
  LISA & $5.95\times10^{-53}$ ($1.24\times10^{-52}$) &
$1.20\times10^{-35}$ ($3.75\times10^{-35}$) & $4.99\times10^{-10}$
($2.07\times10^{-9}$) & $6.66\times10^{-2}$ ($2.36\times10^{-1}$) \\
 ET & $5.08\times10^{-48}$ & $1.41\times10^{-34}$ & $1.52\times10^{-14}$ & $2.63\times10^{-8}$ \\
\hline
 \end{tabular}
\caption{Median of upper bounds on $\mathbb{A}$ (in units of ${\rm
eV}^{2-\alpha}$) obtained over a range of masses for
advanced LIGO, the Einstein Telescope, Cosmic Explorer, and LISA sensitivities, which represent
second-generation, third-generation, and space-based detectors. The bounds
quoted are for nonspinning systems, while the ones in brackets are
bounds from systems with an effective spin $\chi=0.4$. The sources for
the ground-based detectors are assumed to be at a redshift of 0.2, while
those for LISA are assumed to be at a redshift of 0.5.  
See Figs. \ref{fig:A-spin-alpha-lt2} and \ref{fig:A-spin-alpha-gt2} for details.
}
\label{tab:summary}
\end{table*}
\bibliographystyle{apsrev}
\bibliography{LIV}

\end{document}